# ENERGY HARVESTING BY OPTIMIZED PIEZO TRANSDUCTION MECHANISM

**Bijo Boban, S. K. Suresh Babu, U. Satheesh, D. Devaprakasam***
NEMS/MEMS/NANOLITHOGRAPHY Lab,
Department of Nanosciences and Technology, Karunya University, Coimbatore-641114, India
Email: devaprakasam@karunya.edu

**Abstract:** We report generation of electrical energy from nonlinear mechanical noises available in the ambient environment using optimized piezo transduction mechanisms. Obtaining energy from an ambient vibration has been attractive for remotely installed standalone microsystems and devices. The mechanical noises in the ambient environment can be converted to electrical energy by a piezo strip based on the principle of piezoelectric effect. In this work, we have designed and developed a standalone energy harvesting module based on piezo transduction mechanisms. Using this designed module we harvested noise energy and stored electrical energy in a capacitor. Using NI-PXI workstation with a LabVIEW programming, the output voltage of the piezo strip and voltage of the capacitor were measured and monitored. In this paper we discuss about the design, development, implementation, performance and characteristics of the energy harvesting module.

## 1. INTRODUCTION

Recent advancement in the fabrication technology is effectively utilized for design and development of energy efficient microelectronic devices, standalone communication systems and micro electro mechanical systems (MEMS) [3, 4, 6, 7]. The harvested energy from the ambient environment can been used to power the standalone micro systems and devices which are installed in remote locations [1, 2]. The piezo transducer converts the noises of mechanical vibrations energy in to the electrical energy and the converted energy are stored in the capacitor of EH module [7]. The purpose of using energy harvesting system is to power the standalone and handheld devices and recharge and replenish the consumed energy without external intervention [14].

In this paper we report design, development, implementation and characterization of piezoelectric based energy harvester module which harvest the energy noise energy from the mechanical vibrations present in the ambient environment. When mechanical noises vibrate piezo transducer which generates the voltage signal and converted energy is stored in the capacitor. The piezoelectric source produces power in the rate of few mill-watts [5]. This power is accumulated in the harvesting module and stored in capacitor.

We used Orcad simulation tool for designing and simulating of energy harvesting module circuit. We designed energy harvesting PCB by using Advanced Linear Devices (ALD). LabVIEW software used to interlink the module with the NI-PXI workstation, which is used to measure and store the harvested data in PC for further analysis.

## 2. CIRCUIT DESRIPTION

The circuit diagram of the proposed design of energy harvesting module is shown in below Figure 1. The circuit consisted of bridge rectifiers, harvesting device and PIC controller. This was been designed and tested in the Orcad simulation tool. The piezo electric strip is connected to the bridge rectifier and then it is connected to the energy harvesting module which is connected to capacitive device [15].

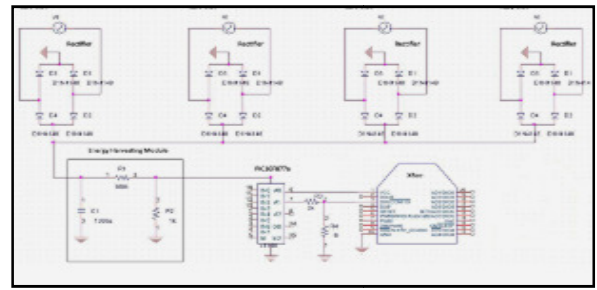

Figure 1. Shows circuit diagram of energy harvesting model implemented in Orcad software tool.

The energy harvesting circuit continuously collects the energy obtains from the piezo strips and it is stored in 3300µf/5V capacitor.

## 3. EXPERIMENTAL PROCDEURE

The piezo strip cantilever was kept in the ambient environment. The noises in the ambient environment make the piezo strip to vibrate due to piezo transaction mechanism, the voltage is generated which is stored in the harvesting module [4, 8, 9]. From a single piezo strip we obtained maximum of 3 volt. The results show (figure 5 & 6) that the generated voltage reaches 3V in few minutes. Both generated and stored voltages were measured in LabVIEW software and the data of the real time measurement for monitoring and analyzing of harvesting system [10, 11].

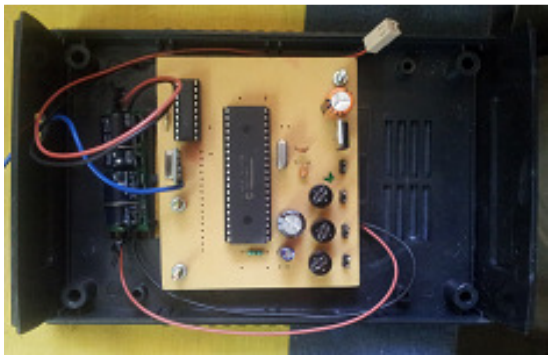

Figure 2. Shows the hardware part of energy harvested module and the designed PCB board.

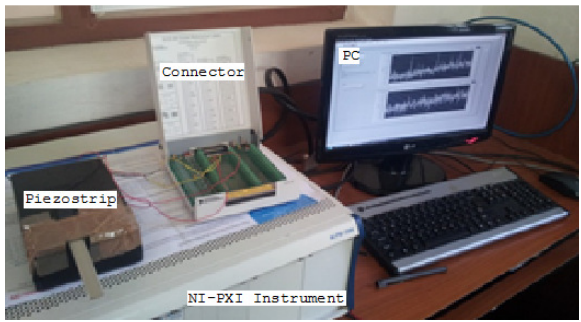

Figure 3. The energy harvesting and measuring experimental setup with NI-PXI workstation.

## 4. RESULTS AND DISCUSSION

The harvested energy from the environment is measured using NI-PXI workstation and LabVIEW software. Figure 4 shows Labview block diagram design layout for measuring the voltages of piezo strip and harvester voltage

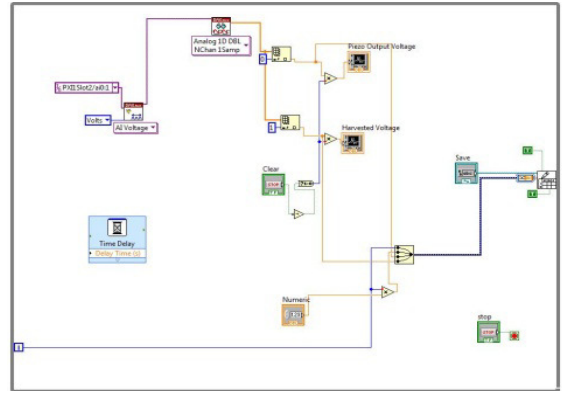

Figure 4: LabVIEW circuit layout for measuring harvested energy

Using LabVIEW the voltage across piezo strip and the energy harvested circuit were been acquired and displayed in Figure 5. From this analysis normal environment it can harvests noises and converts into electrical energy and output voltage signal reaches around 3 V. It shows energy obtained in the normal environment without any disturbance.

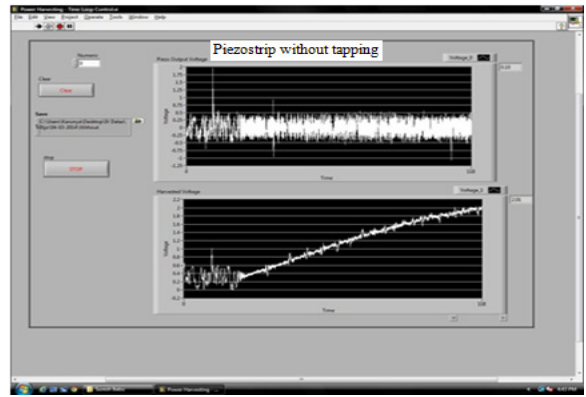

Figure 5. Screen shot of data acquisition in LabVIEW under the ambient condition. Signal from piezo strip and increase of voltage in the storage capacitor.

In another condition the piezo strip were disturbed/tapped at three intervals, it shows energy harvested rapidly at these points [12,13].

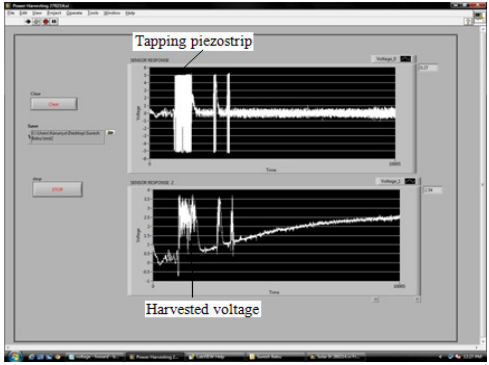

Figure 6. Screen shot of data acquisition in LabVIEW under the external perturbation and ambient condition. signal from piezo strip and increase of voltage in the storage capacitor

## 5. POWER SPECTRAL DENSITY ANALYSIS

The power spectral density (PSD) of the signal, describes the power contributed to the wave, by a frequency, per unit frequency. Power spectral density is commonly expressed in watts per hertz. We analyzed the output voltage signal of the piezo strip using various PSD algorithms (Thompson Multiaper PSD, Periodogram PSD) available in the Matlab program. This analyzed result is shown in Figure 8.

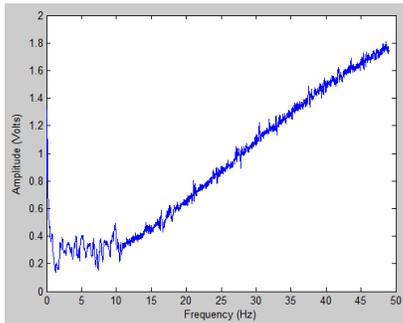

Figure 7. Shows the output voltage signal of the piezo strip.

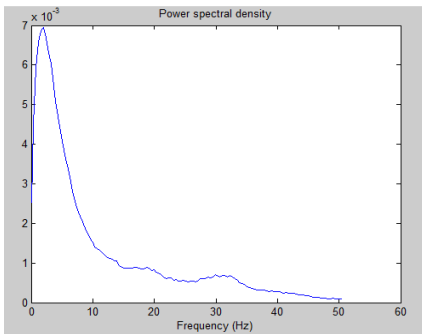

Figure 8. The PSD response of the voltage signal from harvesting circuit under the ambient condition.

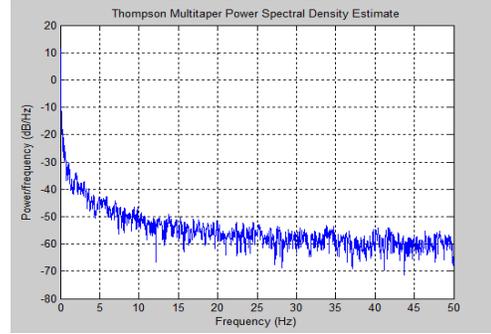

Figure 9. The Thompson Multiaper PSD estimates response of the signal from harvesting circuit

The thompson multitaper method is used to reduces estimation bias by obtaining multiple independent estimates from the same sample insisted of averaging the signal. By this estimation method the attenuation loss of signal is eliminated. This obtain result is shown in figure 9.

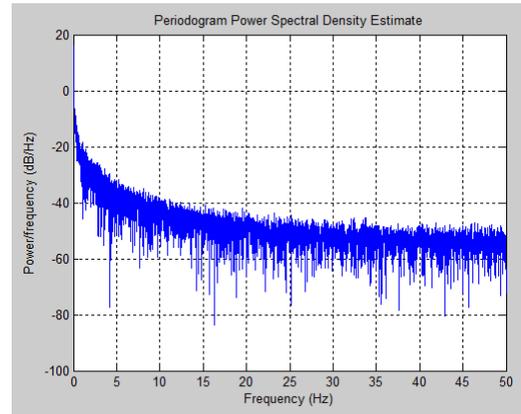

Figure 10. The Periodogram PSD estimates response of the harvested signal from harvesting circuit, under the ambient condition.

Periodogram is used to estimate the total power of the harvested signal, In order to conserve this total power, we multiply all the frequencies that occur in both sets of positive and negative frequencies. This estimated result is shown in figure 10.

## 6 .CONCLUSION & FUTHER WORK

The results of our studies show that the noise energy in ambient environment is successfully harvested and stored by the designed energy harvesting module. The designed EH module can be effectively utilized in many demanding applications because of its its flexibility and low cost. The EH can be used to power small standalone electronic devices installed in remote locations.

Further research work will be carried out with array of piezo strips to obtain to harvest maximum energy and improve its efficiency for the remotely installed small standalone applications. PSD analysis shows the high intensity of noises available in the ambient environment of low frequencies and this could be used for maximize the efficiency of energy harvesting.

ACKNOWLEDGEMENT

We thank DST-Nanomission, Government of India and Karunya University for providing the financial support to carry out the research. We also thank the department of Nanosciences and technology for the help and support to this research.